\documentclass[11pt]{llncs}

\newtheorem{fact}{Fact}
\newtheorem{method}{Method}
\usepackage{latexsym}
\newcommand{\dom}{\{0,1\}^n\times\{0,1\}^n }
\newcommand{\fdom}{\{0,1\}^n\times\{0,1\}^n\to\{0,1\} }
\begin{document}
\title{Rectangle Size Bounds and Threshold Covers in Communication Complexity}
\author{Hartmut Klauck\thanks{Supported by the EU 5th framework program QAIP IST-1999-11234
  and by NWO grant 612.055.001.}}
\institute{CWI\\P.O.-Box 94079\\
1090GB  Amsterdam,  The Netherlands\\
{\tt klauck@cwi.nl}}
\maketitle

\begin{abstract}
We investigate the power of the most important lower bound technique
in randomized communication complexity, which is based on
an evaluation of the maximal size of approximately monochromatic
rectangles, minimized over all distributions on the inputs.
While it is known that the 0-error version of this bound
is polynomially tight for deterministic communication, nothing
in this direction is known for constant error and randomized
communication complexity. We first study a
one-sided version of this bound and obtain that its value lies
between the $MA$- and $AM$-complexities of the considered function. Hence
the lower bound actually works for a (communication
complexity) class between $MA\cap co-MA$ and
$AM\cap co-AM$. We also show that the $MA$-complexity of the
disjointness problem is $\Omega(\sqrt{n})$. Following this we consider
the conjecture that the lower bound method is polynomially tight for randomized
communication complexity. First we disprove a distributional version
of this conjecture. Then we give a combinatorial characterization of
the value of the lower bound method,
in which the optimization over all distributions is absent. This
characterization is done by what we call a uniform threshold cover.
We also study relaxations of this notion, namely approximate majority covers and majority covers,
and compare these three notions in power, exhibiting exponential
separations. Each of these covers captures a lower bound method
previously used for randomized communication complexity.
\end{abstract}

\section{Introduction}
Communication complexity has grown into a central area in theoretical
computer science since the seminal article by Yao \cite{Y79}, finding
more and more applications that range from from VLSI
resource-tradeoffs (e.g.~\cite{T79}) to data-stream
computations (e.g.~\cite{SS02}), see the excellent monography \cite{KN97} for pre-1997
applications. While communication complexity has been helpful by
inspiring upper bounds in other models, its main importance lies in
the lower bounds it provides. Often also variations of the techniques first devised
for communication complexity become important in other areas, e.g.~the
field of branching programs (see \cite{W00}).

Lower bounds for deterministic communication complexity are
usually not very hard to prove, but they are often not strong enough in
applications. Considering randomized communication complexity is
frequently necessary. The lower bound on the monotone circuit
depth of the matching function given in \cite{RW92} is an example where
randomized communication complexity is used to prove a lower
bound for some resource in a deterministic model. Furthermore, since
randomized algorithms are considered standard today and any problem
for which we can describe an efficient randomized algorithm is
considered tractable, lower bounds on the randomized communication
complexity are necessary to show that a communication problem is
hard. Also, communication complexity is an interesting scenario to study
the power of randomization.

Basically all lower bounds on randomized communication complexity with
bounded error are derived by considering properties of rectangles in
the communication matrix. These proofs are usually done in two steps. First,
so-called distributional communication complexity is
considered. The distributional deterministic communication complexity with error $\epsilon$ under
a distribution $\mu$ on the inputs
is the minimal complexity of a deterministic protocol computing a function
while erring with probability at most $\epsilon$ under $\mu$.
According to the Yao-principle the randomized
complexity of a problem equals the maximum over all
distributions of the distributional deterministic complexity. Hence this first step
can always be done without loss of generality (or degradation of the
bounds).

After choosing an appropriate distribution on the inputs one is left
to analyze the deterministic distributional communication complexity.
The $2^c$ message sequences used by a communication $c$ protocol
partition the communication matrix into $2^c$ rectangles\footnote{For
 a definition of communication matrices and rectangles see
 Definition~\ref{def:ccm}.} labeled with the output of the protocol
on that message sequence. Proving a
lower bound on the number of rectangles needed in such a partition is
then done by showing that all $1-\epsilon$-correct rectangles are small.
This approach or variants of it have been used by Yao \cite{Y83}, Babai et
al.~\cite{BFS86}, Razborov \cite{R92}, and adapted to partial
functions also by Raz \cite{R99}, so that virtually all important
lower bound proofs (except \cite{KS92}) for randomized communication
complexity follow the described pattern.

More precisely, the lower bound method (as described by Yao \cite{Y83}) goes as follows:
First one fixes a distribution that puts roughly as much weight on the
1-inputs as on the 0-inputs of a function $f$. One decides whether one proves a
bound on the size of rectangles containing predominantly 1-inputs or 0-inputs.
Then one shows that all rectangles of the desired type with size larger than $1/2^k$ must
contain an $\epsilon$-fraction of wrongly classified inputs. As the
consequence the
randomized communication complexity of $f$ is $\Omega(k)$.

The main question motivating this paper is whether this lower bound
method is tight, i.e., whether we may {\it always} prove lower bounds at most
polynomially smaller than the actual randomized communication
complexity using this method.
This question is stated as Open Problem 3.23 in \cite{KN97}. Since the method
already yields lower bounds, answering the question in the affirmative
demands showing an upper bound on
the randomized communication complexity in terms of the maximal value
obtained by the lower bound method.

It is well known that this is possible in the case $\epsilon=0$,
i.e., the corresponding lower bound method for deterministic protocols
based on the size of monochromatic rectangles always yields results
being at most quadratically
smaller that the deterministic communication complexity. This result
can be proved in two steps: first the 0-error rectangle bound is
characterized via
nondeterministic communication complexity (see Theorem 2.16 in \cite{KN97}).
Then the deterministic communication complexity is upper bounded by the product of
the nondeterministic and co-nondeterministic communication
complexities \cite{AUY83}, Theorem 2.11 in \cite{KN97}.
In this paper we consider the analogous question in the situation when
the error probability is larger than 0.

Note that all proofs in this paper are provided
in the appendix.

\section{Power of the rectangle bound}

First let us fix some notation and give a formal definition of the
main lower bound method investigated in this paper.

\begin{definition}
Let $\mu$ be a distribution on $\dom$ and $\alpha\le1/2$.
Then $\mu$ is $\alpha$-balanced for $f:\fdom$, if \[\alpha\le \mu(f^{-1}(1)),\mu(f^{-1}(0))\le
1-\alpha.\]

$1/2$-balanced distributions are called strictly balanced,
$1/4$-balanced distributions are called balanced.
\end{definition}

\begin{definition}\label{def:ccm}
The communication matrix of a function $f:\fdom$ is a matrix $M_f$ with
rows and columns each corresponding to
$\{0,1\}^n$, and with $M_f(x,y)=f(x,y)$.

A rectangle is a product set in $\dom$.
Rectangles are labeled, a $v$-rectangle being
labeled with $v\in\{0,1\}$. $v(R)$ gives the label of $R$.

The size of a rectangle (or any other set) $R$ with regard to some distribution $\mu$ on
$\dom$ is $\mu(R)=\sum_{x,y\in R} \mu(x,y)$. Let
$err(R,\mu,v)=\mu(f^{-1}(1-v)|R)$ denote the error of a
$v$-rectangle $R$.
\end{definition}

We consider both one-sided and two-sided versions of the rectangle bound.

\begin{definition}
\[size(\mu,\epsilon,f,v)=\max\{\mu(R)\,:\,err(R,\mu,v)\le
\epsilon\}, \]
where $R$ runs over all rectangles in $M_f$.

\[bound^{(1)}_\epsilon(f)=\max_\mu\log(1/size(\mu,\epsilon,f,1)),\]
where $\mu$ runs over all balanced distribution on $\dom$.

Furthermore
\[bound_\epsilon(f)=\max\{bound^{(1)}_\epsilon(f),bound^{(1)}_\epsilon(\neg f)\}.\]

We use the conventions
$bound(f)=bound_{1/4}(f)$ and $bound^{(1)}(f)=bound^{(1)}_{1/4}(f)$.\end{definition}

Let us first note a fundamental property of the rectangle bound,
namely error reducibility.

\begin{lemma}\label{lem:err-red}
Let $\epsilon\le 1/2-\Omega(1)$.

Assume $bound^{(1)}_\epsilon(f)=k$. Then
$bound^{(1)}_{\epsilon^l}(f)\le O(lk)$.

Assume $bound_\epsilon(f)=k$. Then $bound_{\epsilon^l}(f)\le O(lk)$.
\end{lemma}

The lemma is proved in appendix C. Also note that the definition is
almost invariant with respect to the ``balancedness'' of the
underlying distribution, see again appendix C for a proof.

\begin{lemma}\label{lem:bal}
Assume 
$\max_\mu\log(1/size(\mu,\epsilon,f,1))=k$ where $\mu$ runs over all
$\alpha$-balanced distribution on $\dom$ for some constant $\alpha$. Then
\[k=\Theta(bound^{(1)}_{\epsilon}(f)),\] given that $\epsilon\le \alpha/4$.
\end{lemma}

For definitions of the different communication complexity modes
considered in this paper see appendix B. Note that all randomized
modes of communication complexity are defined to have a public coin there.

It is well known that $bound(f)$ yields a lower bound on the
randomized communication complexity of $f$, see \cite{KN97}. So we
have the following lower bound method.

\begin{method}[$\epsilon$-error randomized communication complexity]
\ \begin{enumerate}
\item Pick a balanced distribution on $\{0,1\}^n\times\{0,1\}^n$.
\item Pick $v\in\{0,1\}$.
\item Show that all $1-\epsilon$-correct $v$-rectangles in $M_f$ have size
 $<2^{-b}$.
\item Then $R_\epsilon(f)\ge \Omega(b)$.
\end{enumerate}
\end{method}

As an example we give the following fact due to Razborov \cite{R92},
which will be used several times in this paper.
Let $DISJ(x,y)=\wedge_{i=1}^n (\neg x_i\vee\neg y_i)$ be the set
disjointness problem. 

\begin{fact}\label{fac:Razb}
For $DISJ$ there is an balanced distribution $\mu$ on $\dom$, so
that every 0-rectangle $R$ in $\dom$ either satisfies $\mu(R)\le
2^{-\beta n}$ or $err(R,\mu,0)\ge\epsilon$ for some constants
$\beta,\epsilon>0$.

In other words (using Lemma~\ref{lem:err-red}), $R(DISJ)\ge bound^{(1)}(DISJ)=\Omega(n)$.
\end{fact}

We now try to determine exactly for which class of problems the
lower bound method works. We show that $bound^{(1)}(f)$ lies between the $MA$- and
$AM$-complexities of $f$.

\begin{theorem}\label{the:AMMABD}
\begin{enumerate}
\item For $f:\fdom$ and $\epsilon\in[1/2^{MA(f)},1/2-\Omega(1)]$:
\[MA_\epsilon(f)\ge\Omega\left(\sqrt{bound_\epsilon^{(1)}(f)}\right).\]
\item For all $f:\fdom$ and $\epsilon\le 1/2-\Omega(1)$:
\[AM_\epsilon(f)\le O(bound_\epsilon^{(1)}(f)+\log(1/\epsilon)).\]
\end{enumerate}
\end{theorem}

With this theorem and Fact~\ref{fac:Razb} we can 
conclude a new lower bound in communication complexity. 

\begin{corollary}\label{cor:MADISJ}
\[MA(DISJ)=\Omega(\sqrt{n}),\]
while \[N(\neg DISJ)=O(\log n).\]
\end{corollary}

It seems unlikely that, but remains unknown whether $DISJ$ has efficient
$AM$-protocols. Actually no separation between larger classes than
$MA\neq co-MA$ is known within
the communication complexity version of the polynomial hierarchy
(see \cite{BFS86}, polynomial time is replaced by
polylogarithmic communication in this definition).
Note that it is still open
whether the polynomial hierarchy in communication complexity is
strict. Actually we will give a lower bound in
Theorem~\ref{the:APPPP} showing that some explicit function is not contained in
some even larger subclass of the polynomial hierarchy in communication
complexity than $MA\cup
co-MA$, yet that function itself probably is not included in the
hierarchy, as opposed to $DISJ$.

We can conclude the following relations between Arthur Merlin and
randomized communication and the lower bound method.

\begin{corollary}\label{cor:classes} 
\[R(f)\ge\Omega(bound(f)),\]
\[R(f)\ge\Omega(\max\{MA(f),MA(\neg f)\})\ge
\Omega\left(\sqrt{bound(f)}\right),\]
\[bound(f)\ge\Omega(\max\{AM(f),AM(\neg f)\}).\]
\end{corollary}

If we could show that any function with both small $AM$-
and $co-AM$-complexity also has small randomized complexity we could
show that lower bound method 1 is always polynomially tight. If, on
the other hand, $R(f)\ge g(bound(f))$ for some superpolynomial $g$ and
some $f$, then there is a superpolynomial separation between
$\max\{AM(f),AM(\neg f)\}$ and $R(f)$.

The first attempt to prove tightness of $bound(f)$ coming to mind uses the
Yao-principle and switches to (non-) deterministic distributional
complexity with error in the hope to employ
similar techniques as in previous combinatorial
results \cite{AUY83}, where $D(f)\le
O(N(f)\cdot N(\neg f))$ is shown for all $f:\fdom$.

It is well known (see Theorem 3.20 in \cite{KN97}) that
\begin{fact}\label{fac:Rdistr}
\[R_\epsilon(f)=\max_\mu D_\epsilon^\mu(f).\]\end{fact}

We observe that by the same proof
\begin{lemma}\label{lem:AMdistr}
\[AM_\epsilon(f)=\max_\mu N_\epsilon^\mu(f).\]\end{lemma}

Hence if we could relate the $\epsilon$-error distributional nondeterministic
complexity $N_\epsilon^\mu(f)+N_\epsilon^\mu(\neg f)$ to the
$\epsilon$-error distributional deterministic
complexity $D_\epsilon^\mu(f)$ for all distributions $\mu$ we would
have shown that the rectangle bound is
always polynomially tight. But the approach does not work as shown in
the next result.

\begin{theorem}\label{the:distribconj}
There is a function
 $WHICH:\{0,1\}^n\times\{0,1\}^n\to\{0,1\}$, a balanced
distribution $\mu$ on
 $\{0,1\}^n\times\{0,1\}^n$, and a constant $\epsilon>0$ so that
\[N_0^\mu(WHICH),N_0^\mu(\neg WHICH)=O(\log n),\]
\[D_\epsilon^\mu(WHICH)=\Omega(n).\]
\end{theorem}

So this first attempt to prove that the rectangle bound is tight,
fails. Proving $R(f)\le poly(AM(f)+AM(\neg f))$ requires an argument
not considering the distributional complexity for all distributions separately.
Also note that the distributions maximizing
$N_\epsilon^\mu(f),N_\epsilon^\mu(\neg
f),D_\epsilon^\mu(f)$ are in general not the same.

The proof of Theorem~\ref{the:distribconj} establishes a lower bound
on the number of rectangles needed to partition (with small error) the communication
matrix into rectangles, while
errorfree covers (with overlapping rectangles), and hence large
errorfree 1- and
0-rectangles exist.

\section{The rectangle bound and bounded error uniform threshold covers}

In this section we start another approach to prove that the lower bound
method is tight. Instead of considering Arthur Merlin complexity we
characterize the lower bound method itself combinatorially.

\begin{definition}
A uniform threshold cover with parameters $s,t$ for a communication
problem $f:\fdom$ is a set of rectangles in the communication matrix
of $f$ with labels from $\{0,1\}$, so that for each
input $x,y$ at least $t$ of the adjacent rectangles bear the correct
label $f(x,y)$ and at most $s$ of the adjacent rectangles bear the
wrong label $1-f(x,y)$.

A one-sided uniform threshold cover is as above, but only 1-labeled
rectangles are used, and inputs with $f(x,y)=1$ lie in at least $t$
rectangles, while inputs with $f(x,y)=0$ lie in at most $s$ rectangles.

Let $f:\fdom$ be a communication problem. Let $P$ denote the
minimal size of a one-sided uniform threshold cover with parameters
$s,t$ for $f$. Then
$UT^{(1)}_{s,t}(f)=\lceil \log P\rceil$ is called the one-sided
uniform threshold complexity of $f$.

Let $f:\fdom$ be a communication problem. Then
$UT_{s,t}(f)=\max\{UT^{(1)}_{s,t}(f),UT^{(1)}_{s,t}(\neg f)\}$ is called the
uniform threshold complexity of $f$, and equals (within $\pm 1$) the logarithm of the
minimal size of a uniform threshold cover for $f$.

We will say that a uniform threshold cover with parameters $s,t$ has bounded error, if $t\ge 2s$.
\end{definition}

The main features of bounded error uniform threshold covers are first, that the
acceptance threshold is the same for {\it all} inputs, and secondly,
the bounded error.

\begin{remark}\label{rem:boostUT}
Given a bounded error uniform threshold cover for $f$ of size $2^k$
with the parameters $s,2s$, we can form all possible $l=\log(1/\epsilon)$ tuples of
1-rectangles, and take the intersections of the
rectangles in such tuples into a new cover, labeled as
1-rectangles. Then we proceed analogously with the 0-rectangles. Clearly each input is in
$(2s)^l=(1/\epsilon)\cdot s^l$ correctly labeled rectangles, and in at
most $s^l$ incorrectly labeled rectangles. Hence, there is a value
$s'=(1/\epsilon)s^l$ with $UT_{\epsilon s',s'}(f)\le O(k\cdot \log(1/\epsilon))$.
\end{remark}

We now characterize the lower bound method in terms of bounded
error uniform threshold covers.

\begin{theorem}\label{the:BDUT}
\begin{enumerate}
\item \begin{enumerate}
\item $bound^{(1)} (f)\le O(UT^{(1)}_{s,2s}(f))$.
\item $bound (f)\le O(UT_{s,2s}(f))$.
\end{enumerate}
\item \begin{enumerate}
\item $UT^{(1)}_{n,n^2}(f)\le O(bound^{(1)} (f)\cdot\log n)$.
\item $UT_{n,n^2}(f)\le O(bound (f)\cdot\log n)$.
\end{enumerate}\end{enumerate}
\end{theorem}

Note that $UT_{s,2s}^{(1)}(f)\le O(bound (f))$ is not always true,
as we will show after Theorem~\ref{the:APPUT} in
Remark~\ref{rem:tight}.

So we have a quite natural version of covers that captures the
technique used in most of the
lower bounds for randomized communication complexity. Showing that
$R(f)\le poly(UT_{n,n^2}(f))$ for all $f:\fdom$ would immediately show
tightness of $bound(f)$. We have been unable to prove such a result so
far. Nevertheless, Theorem~\ref{the:BDUT} turns the problem of showing tightness
of the lower bound method into a combinatorial one, not
involving a maximization over distributions. Alternatively we may
reformulate the problem as follows.

\begin{corollary}\label{cor:probl}
Let $UT[R]: R\to \{0,1\}$ for a rectangle $R\subseteq
\{0,1\}^N\times\{0,1\}^N$
be the following communication problem (for some value $t$ depending
on $R$):
\[
UT[R](x,y)=
\left\{\begin{array}{lll}
1&\mbox{ if } \sum_{i=1}^{N/2} x_i\wedge y_i\ge t&\mbox{ and }
\sum_{i=N/2+1}^{N} x_i\wedge y_i\le \sqrt{t}\\
0&\mbox{ if } \sum_{i=1}^{N/2} x_i\wedge y_i\le \sqrt{t}&\mbox{ and }
\sum_{i=N/2+1}^{N} x_i\wedge y_i\ge t\\
\mbox{undef}&\mbox{ else.}
\end{array}\right.
\]
A protocol for computing $UT[R]$ works under the promise that $R$ contains
only defined inputs.
A protocol computes $UT$, iff for each $R$ (that contains only defined
inputs) the players (knowing $R$) compute
$UT[R]$ correctly.

\begin{eqnarray*}
\mbox{Then }&
R(UT)\le poly(\log N)\\
&\iff \forall f:\fdom: R(f)\le poly(bound(f)\log n).
\end{eqnarray*}
\end{corollary}

Note that for the communication problem $UT$, the threshold $t$ is
usually much smaller than $N$.

So it is sufficient (and necessary) to give an efficient randomized protocol for the
promise problem $UT$ to show tightness of $bound(f)$.

\section{Comparing different notions of threshold covers}

We now
consider variations of the notion of threshold covers. The
most immediate is a majority cover.

\begin{definition}
A majority cover for a function $f$ is a set of labeled rectangles so
that for each input the majority of the adjacent rectangles bears the
correct label. Ties are broken in favor of $f(x,y)=1$.

Let $PP(f)$ denote the logarithm of the size of a smallest majority
cover for a function $f$.
\end{definition}

The above notion of majority covers corresponds to majority
nondeterministic protocols, which accept an input,
whenever there are more nondeterministic computations leading to
acceptance than to rejection: each computation in a
nondeterministic protocol corresponds to a rectangle.
Majority covers are also equivalent to randomized protocols with
error ``moderately'' bounded away from 1/2 as shown in \cite{HR90}.

\begin{fact}\label{fac:HR90}
There is a majority cover of size $2^{k}$ for $f:\fdom$, iff
there are $\epsilon,c$ with $c+\log(1/\epsilon)=\Theta(k)$, as well as
a randomized protocol with error $1/2-\epsilon>0$ and
communication $c$ computing $f$ (in the protocol the players are allowed to use a
private source of randomness only).
\end{fact}

Note that a randomized protocol can be viewed as a probability
distribution on deterministic protocols, and that each deterministic
protocol induces a partition of the communication matrix into
rectangles labeled with the function value. Then the union of all
these rectangles is a 
uniform threshold cover for $f$, though not one with bounded error. The
number of rectangles used is this cover is upper bounded by the number
of message sequences used in the randomized protocol (here we use the
fact that the randomized protocol can access private random sources only).

\begin{corollary}
$PP(f)\le O(k)\iff \exists s:UT_{s,s+1}(f)\le O(k)$.
\end{corollary}

Hence, if we drop the bounded error feature from uniform threshold
covers, we can as well drop the uniformity feature.

It is shown in \cite{Kl01} that majority covers have a
strong connection to a lower bound method in communication complexity
based on discrepancy.

\begin{definition}
Let $\mu$ be any distribution on
 $\{0,1\}^n\times\{0,1\}^n$ and $f$ be any function
 $f:\{0,1\}^n\times\{0,1\}^n\to\{0,1\}$.
Then let \[disc_\mu(f)=\max_R|\mu(R\cap f^{-1}(0))-\mu(R\cap f^{-1}(1))|,\]
where $R$ runs over all rectangles in $M_f$.
Denote $disc(f)=\min_\mu disc_\mu (f).$
\end{definition}

\begin{fact}\label{fac:PPdisc}
For all $f:\fdom$:
\[\log(1/disc(f))\le PP(f)\le O(\log 1/disc(f)+\log n).\]
\end{fact}

Note that discrepancy $1/2^k$ under some distribution essentially
means that all rectangles with size at least $1/2^{k/2}$ have error at
least $1/2-1/2^{k/2}$.
We can prove lower bounds for the $PP$-complexity in the following way.

\begin{method}[$PP$-complexity, discrepancy]
\ \begin{enumerate}
\item Pick a distribution $\mu$ on $\{0,1\}^n\times\{0,1\}^n$.
\item Show that all rectangles $R$ in $M_f$ have $|\mu(R\cap
  f^{-1}(0))-\mu(R\cap f^{-1}(1))| <2^{-b}$.
\item Then $PP(f)\ge \Omega(b)$.
\end{enumerate}
\end{method}

Consider the function $MAJ:\fdom$ with \[MAJ(x,y)=1\iff \sum_{i=1}^n (x_i\wedge
y_i)\ge n/2.\] Obviously $MAJ$ has a majority cover of size $O(n)$. It is
easy to see that $DISJ$ and its
complement can both be reduced to $MAJ$. Hence we can easily separate
majority covers from one-sided bounded error uniform threshold covers using
Fact~\ref{fac:Razb}: $bound^{(1)}(DISJ)=\Omega(n)$ which implies with
Theorem~\ref{the:BDUT} that $UT^{(1)}_{s,2s}(DISJ)=\Omega(n)$.

\begin{corollary}\label{lem:PPUT}
$PP(MAJ)=O(\log n).$

$UT^{(1)}_{s,2s}(MAJ)=\Omega(n)$ all $s$.
\end{corollary}

So a majority cover is in fact much stronger than even a one-sided
uniform threshold cover with bounded error.
Let us now consider a relaxation of majority covers that has bounded error
in some sense. Compared to bounded error uniform threshold covers we now
drop the uniformity constraint on the threshold.

\begin{definition}
An approximate majority cover is a majority cover in which for each input at
least $3/4$ of the adjacent rectangles bear the correct label. Let
$APP(f)$ denote the logarithm of the size of a minimal approximate majority
cover for $f$.
\end{definition}

\begin{remark}\label{rem:APPboost}
Note that the parameters $1/4, 3/4$ can be improved to arbitrary
constant $\epsilon,1-\epsilon$ by forming $k$-tuples of rectangles and
taking their intersections as the new approximate majority cover with $k=\log(1/\epsilon)$.
\end{remark}

The definition of approximate majority covers is similar to threshold
computations on Turing machines in a class named $BPP_{path}$ as
considered in \cite{HHT97}. We
prefer our naming to $BPP_{path}$, since the class has little
similarity to $BPP$ and is not defined in terms of paths here.
It is shown in \cite{HHT97} that $BPP_{path}$ contains $MA\cup
co-MA$ and is hence probably much more powerful than $BPP$.
We immediately get a similar result for communication complexity using
Theorems~\ref{the:AMMABD} and~\ref{the:BDUT}.

\begin{theorem}\label{the:APPUT}
$APP(f)\le O(UT^{(1)}_{s,2s}(f))$ for all $s$.

$APP(f)\le O(UT^{(1)}_{s,2s}(\neg f))$ for all $s$.

$APP(f)\le \min\{O(MA(f)^2), O(MA(\neg f)^2)\}$.
\end{theorem}

\begin{remark}\label{rem:tight}
It is easy to see that $PP(EQ)=\Theta(\log n)$ for the equality
function $EQ(x,y)=1\iff x=y$, since $PP(f)\ge\log D(f)$. Hence
$UT^{(1)}_{s,2s}(EQ)=\Omega(\log n)$. On the other hand
$bound(EQ)=\Theta(1)$, since $R(EQ)=\Theta(1)$ (see Example 3.13 in
\cite{KN97} and note that randomized
protocols are defined to have a public coin here). Hence the relation
$UT^{(1)}_{s,2s}(f)\le O(bound(f)\log n)$ from
Theorem~\ref{the:BDUT} cannot be improved to $\le O(bound(f))$, but
possibly to $\le O(bound(f)+\log n)$.
\end{remark}

$APP$-complexity has an interesting connection to a lower bound method
as follows.

\begin{theorem}\label{the:APPrect}
If $APP(f)=k$ then for all balanced distributions $\mu$ there is a
rectangle of size $1/2^{O(k)}$ with error 1/4.

If for all balanced distributions $\mu$ there is a rectangle of size
$1/2^k$ and error $1/4$ then $APP(f)\le O(k)+\log n$.
\end{theorem}

Thus given that $APP(f)$ is small, there is a large rectangle with
small error for each distribution, sometimes a 1-rectangle, sometimes
a 0-rectangle.
We are lead to the following lower bound method.\\

\begin{method}[$APP$-complexity]
\ \begin{enumerate}
\item Pick a balanced distribution on $\{0,1\}^n\times\{0,1\}^n$.
\item Show that all $1-\epsilon$-correct rectangles in $M_f$ have size
  $<2^{-b}$.
\item Then $APP(f)\ge \Omega(b)$.
\end{enumerate}
\end{method}
Actually it has been shown by Yao in \cite{Y83} that for some explicit function 
and some balanced distribution neither large $1-\epsilon$-correct
0-rectangles nor large $1-\epsilon$-correct
1-rectangles exist, hence he demonstrated that this function has
linear $APP$ complexity, which is a much stronger result than his
conclusion that the function has linear
randomized bounded error communication complexity.

We give a separation result between the two types of covers,
stating that approximate majority covers are actually much more powerful than
one-sided bounded error uniform threshold covers and hence
also than $MA$-protocols.

\begin{theorem}\label{the:APPUTsep}
There is a function $BOTH:\{0,1\}^n\times\{0,1\}^n\to\{0,1\}$ so that
\[APP(BOTH)=O(\log n),\]

\[bound^{(1)}(BOTH)\ge\Omega(n);\;\;bound^{(1)}(\neg BOTH)\ge\Omega(n).\]
Hence also
\[UT^{(1)}_{s,2s}(BOTH)\ge\Omega(n);\;\;UT^{(1)}_{s,2s}(\neg BOTH)\ge\Omega(n),\]
\[MA(BOTH)\ge\Omega(\sqrt n);\;MA(\neg BOTH)\ge\Omega(\sqrt{n}).\]
\end{theorem}

So approximate majority covers are exponentially more powerful than
one-sided bounded error uniform threshold covers for $BOTH$ and its complement.
In terms of the lower bound methods this means that for $BOTH$
it is true that for every balanced distribution there is a rectangle of size
$1/poly(n)$ with constant error, but there exists a balanced distribution,
where all 1-rectangles either have error $1/2-o(1)$ or size
$1/2^{\Omega(n)}$, and there exists a balanced distribution,
where all 0-rectangles either have error $1/2-o(1)$ or size
$1/2^{\Omega(n)}$.

To complete the picture we compare the power of $APP$ and $PP$ covers.

\begin{theorem}\label{the:APPPP}
\[PP(MAJ)=O(\log n),\]
\[APP(MAJ)=\Omega(n).\]
\end{theorem}

Note that the above result can be read as saying that for $MAJ$ for all balanced
distributions there exists a rectangle with
discrepancy $1/poly(n)$ (having hence size $1/poly(n)$
and error $1/2-1/poly(n)$), while there is a distribution $\mu$ where
any rectangle with {\it constant} error has
size $1/2^{\Omega(n)}$.

\section{Conclusions}
Virtually all\footnote{A recent exception is a $\Omega(\sqrt{n})$ lower bound on
 the information which must be exchanged in computing
 $DISJ$ \cite{SS02} (and hence on the communication
 complexity). This quantity can also be lower bounded using the
 rectangle method, see \cite{Kl02}.} known lower bounds on randomized
 communication complexity in
the literature can be seen as instances of methods 1, 2, or 3. We have
shown that these three methods have exponential differences in
power. It remains open whether method 1 is polynomially tight for randomized
communication complexity. A way to show this is proving that
the logarithm of the size of bounded error uniform threshold covers is polynomially
related to the randomized communication complexity. This
avoids arguing with the optimum (over all balanced distributions) of the
lower bound parameter. We have shown in Theorem~\ref{the:distribconj} that arguing
for all distributions separately does not yield the desired result.

Methods 2 and 3 have been characterized as more powerful versions of
threshold covers. It is interesting that the rectangle based lower bound proofs can be
understood in terms of these combinatorial objects that are only in
the case of method 2 known to be directly related to standard communication
complexity modes.

We now note some further observations and open problems.
The most significant open problem related to this paper is whether
$UT_{s,2s}(f)$ and $R(f)$ are polynomially
related, resp.~whether $R(UT)\le poly(\log N)$.

It can be shown with techniques as in \cite{HHT97} that every $f$
with $APP(f)=poly(\log n)$ is in the
polylog-communication complexity polynomial hierarchy, see \cite{BFS86} for a
definition of the latter. It is improbable, however, that the same
holds for all functions with $PP(f)=poly(\log n)$, since then the
communication complexity version of the polynomial hierarchy would collapse.
Let us note that the separation of the polynomial hierarchy for
communication complexity is open. Hence method 3 allows to show
that some explicit function is not contained in the class of problems with $APP(f)=poly(\log
n)$, the largest class of problems inside the polynomial hierarchy for
which such a lower bound is known. Showing that this class is a proper subset of
the hierarchy is open. It is also open, which methods might
be applied to separate this hierarchy.

Regarding the fine-structure of the relations between the discussed
complexity measures there are several open problems. Is it possible to
separate $AM$- from $MA$-complexity? This could be done by using the
rectangle method to separate $AM(f)$ from $UT^{(1)}_{s,2s}(f)$. But
it is also possible that $UT_{s,2s}^{(1)}(f)$ is always polynomially related
to $AM(f)$. Furthermore {\it any} lower bounds for $AM$
communication complexity are desirable, since they would probably need new
techniques and lead to progress on the problem of showing lower bounds
for higher classes in the polynomial hierarchy. Also a
separation of $MA(f)$ from $UT^{(1)}_{s,2s}(f)$ would be interesting.

As another issue the role of interaction in communication
complexity is interesting. For nondeterministic
communication 1-round protocols are optimal, not so for
randomized, deterministic, and even communication with limited nondeterminism
\cite{KN97,Kl98}. Clearly 1-round $AM$-protocols are also optimal,
but this seems unlikely for $MA$-protocols. A candidate problem to
establish this conjecture would
be the majority of the outcomes of pointer jumping on $\sqrt{n}$ paths of
length $k$, with the promise that $3/4\cdot \sqrt{n}$ paths lead to the same
output. A randomized protocol with $k$ rounds and $O(k\log n)$
communication can solve this problem, but $MA$-protocols of complexity
$o(\sqrt{n})$ using $k-1$ rounds possibly not.

\clearpage\normalsize

\begin{appendix}

\section{Organization of the rest of the paper}
In appendix B we formally define the different modes of communication
complexity considered in this paper, appendix C provides proofs of
elementary properties of the rectangle bound.
In appendix D we give the proofs concerning the comparison of the
rectangle bound with $MA$- and $AM$-communication complexity, and the
proof of Theorem~\ref{the:distribconj}.
Appendix E shows the equivalence between $bound(f)$ and bounded error
uniform threshold covers for $f$.
Appendix F shows equivalence between lower bound method 3 and approximate majority
covers, and gives separations between the three lower bounds methods
resp.~the three types of threshold covers.

\section{Definitions}

We employ the following definitions of communication complexity.

\begin{definition}
Let $f:\fdom$ be a function. In a communication
protocol players Alice and Bob receive inputs $x,y$ from
$\{0,1\}^n$ each.
Their goal is to compute $f(x,y)$. To this end the players exchange
binary encoded messages. The communication complexity of a
protocol is the worst case number of bits exchanged. 

The deterministic communication complexity $D(f)$ of a function $f$ is
the complexity of an optimal protocol computing $f$.

In a nondeterministic protocol for a Boolean function $f$ the players
are allowed to guess some bits and communicate according to a
different deterministic strategy for each guess string. An input is
accepted iff at least one computation accepts. The nondeterministic
guesses are private and accessible to the guessing player only. The
nondeterministic communication complexity $N(f)$ is the complexity of
an optimal nondeterministic protocol computing $f$.

In a randomized protocol for a Boolean function $f$ the players can
access a public source of random bits.
They can communicate according to a different deterministic strategy for
each value of the random bits. It is required that for each input the
correct output is produced with probability $1-\epsilon$ for some
$\epsilon<1/2$. The randomized communication
complexity $R_\epsilon(f)$ is the complexity of an optimal
randomized protocol computing $f$ with error probability $\epsilon$.

Arthur Merlin computations have been introduced in \cite{B85,BM88}.
In an Arthur Merlin ($AM$) protocol the players may first access a
public source of random bits and read an arbitrarily long random
string. After this phase they start a nondeterministic protocol. A
function $f:\fdom$ is computed if for all $x,y$ with $f(x,y)=1$ with
probability $1-\epsilon$ over the random bits the nondeterministic
protocol accepts (i.e., there is a guess string makes players accept),
while for all $x,y$ with $f(x,y)=0$ with
probability $1-\epsilon$ over the random bits the nondeterministic
protocol does not accept (i.e., there is no guess that makes the
players accept).
The communication complexity of an Arthur Merlin
protocol is the maximum (over the random bits) of the complexities of
the nondeterministic protocols. Let $AM_\epsilon(f)$ denote
the complexity of an optimal Arthur Merlin protocol for $f$ with error
$\epsilon$.

In a Merlin Arthur protocol for a function $f:\fdom$ the players first
make a nondeterministic guess of some length $k$ known to both
players. Then the players perform a randomized protocol. It is
required that for all $x,y$ with $f(x,y)=1$ there is a value of the
guess, so that the protocol accepts with probability $1-\epsilon$,
while for all $x,y$ with $f(x,y)=0$ there is no value of the
guess, so that the protocol accepts with probability larger than $\epsilon$.
The complexity of a Merlin Arthur protocol is given by the maximum complexity
of the communication (over the guesses and the coin tosses) plus $k$. 
Let $MA_\epsilon(f)$ denote
the complexity of an optimal Merlin Arthur protocol for $f$ with error
$\epsilon$.

In case the subscript fixing the error is dropped we set the error to $1/4$.
\end{definition}

Note that an Arthur Merlin protocol is an interactive proof system
with verification performed by a communication protocol. Arthur
challenges Merlin to provide a proof that $f(x,y)=1$,
this proof is verified by Alice and Bob.

A Merlin Arthur protocol uses a randomized protocol to check a fixed proof,
whose length is included in the communication cost. The Merlin Arthur
model would be ill-defined, if we would simply require the
nondeterministic guess to be private and coming without cost. In this case
Alice could simply guess Bob's input nondeterministically, and then use a randomized
protocol for the equality function with $O(1)$ communication to
test if her guess was right (see Example 3.13 in \cite{KN97}, note
that public coin in the randomized protocol). If so, she can compute any function on
$x,y$ and announce the result. Hence any function $f:\fdom$ would have
Merlin Arthur communication complexity $O(1)$ under such a definition.

In lower bound proofs for randomized complexity one often applies the
Yao-principle that states a relation between the complexity in the
randomized setting (with small error probability for every input)
and the complexity in the deterministic setting where correctness is
only demanded with high probability over some distribution on the
inputs.

\begin{definition}
A deterministic protocol has error $\epsilon$ under some distribution
$\mu$ on the inputs, if the probability that the protocol errs is $\epsilon$.

The distributional deterministic complexity of $f:\fdom$ is
$D_\epsilon^\mu(f)$, the minimal complexity of any deterministic
protocol with error $\epsilon$ under $\mu$.

A nondeterministic protocol has error at most $\epsilon$ under some distribution
$\mu$ on the inputs, if the probability (under $\mu)$ of the set of
accepted 0-inputs and nonaccepted 1-inputs is at most $\epsilon$.

The distributional nondeterministic complexity of $f:\fdom$ is
$N_\epsilon^\mu(f)$, the minimal complexity of any nondeterministic
protocol with error $\epsilon$ under $\mu$.
\end{definition}

Nondeterministic communication complexity is related to a specific
type of covers \cite{KN97}.
\begin{fact}
Let $Cov^{(1)}(f)$ denote the minimum number of monochromatic (0-error)
1-rectangles in a set $\{R_1,\ldots, R_c\}$, so that $f^{-1}(1)=\cup_{i=1}^c
R_i\neq\emptyset$.

Then
$N(f)=\lceil \log Cov^{(1)}(f)\rceil$.
\end{fact}

\section{Properties of the rectangle bound}

{\bf Proof of Lemma~\ref{lem:err-red}.}
Assume $bound^{(1)}_\epsilon(f)=k$. Hence for all balanced
distributions there is a $1-\epsilon$-correct 1-rectangle of size
$1/2^k$ at least. Fix any balanced distribution $\mu$. We construct a
rectangle of error $\epsilon^l$ and size $2^{-O(kl)}$ for $\mu$ inductively.

Let $\mu_0=\mu$. First we take a rectangle $R_1$ with error
$\le\epsilon$ and size $s_1\ge1/2^k$
guaranteed by our assumption for $\mu_0$. In case $R_i$ has
no error at all we are done. Otherwise we construct a new
distribution $\mu_1$ as follows: For
all $x,y\not\in R_1$ we set $\mu_1(x,y)=0$. $\mu_1$ is then
normalized to a strictly balanced distribution by multiplying
$\mu_0(x,y)$ by some factor $p_0$ when $f(x,y)=0$ and multiplying $\mu_0(x,y)$
by a factor $p_1$ when $f(x,y)=1$.

Note that since
$\mu(f^{-1}(0)|R)\le\epsilon$ we have $p_0\ge 1/(2\epsilon)$ and
$1/(2(1-\epsilon))\ge p_1\ge 1/2$. Then we can pick a rectangle $R_2$
with error $\epsilon$ and
size $s_2\ge1/2^k$ according to $\mu_1$.

Now we compute the size of $R_2$ according to $\mu$ and also its error
on $\mu$. By concentrating $\mu_1$ on $R_1$ we have increased the
weights of $x,y\in R_1$ uniformly by a factor of $(1/s_1)\le 2^k$. Then we
have balanced the distribution by multiplying 1-inputs' weights with
$p_1$ and 0-inputs' weights by $p_0$.

So the weight of
0-inputs in $R_2$ according to $\mu$ is at most \[s_1\cdot (1/p_0)\cdot \epsilon\cdot
s_2\le s_1\cdot 2\epsilon^2\cdot s_2.\] The weight of the 1-inputs
is at least \[s_1\cdot (1/p_1)\cdot (1-\epsilon)\cdot s_2\ge
s_1\cdot2(1-\epsilon)^2\cdot s_2.\]
The size of $R_2$ is at least $s_1\cdot s_2\ge 1/2^{2k}$.

Assume that $\epsilon\ge 1/4$, then $\epsilon=1/2-\delta$ for some
$\delta\le 1/4$. In this case
\[err(R_2,\mu,1)\le
\frac{\epsilon^2}{(1-\epsilon)^2+\epsilon^2}\le\frac{1}{2}-\frac{\delta}{1/2+2\delta^2}<
1/2-(3/2)\delta.\]
Repeating this $O(1/\delta)=O(1)$ times reduces the error to less than $1/4$.

If $\epsilon\le1/4$, then
$R_2$ has error at most $2\epsilon^2$.
Iterating the construction $O(l)$ times yields the first part of the lemma. Arguing
analogously for 0-rectangles yields the second part.
\qed

{\bf Proof of Lemma~\ref{lem:bal}.}
Let the bound be $k$ when $\mu$ runs over all $\alpha$-balanced
distributions on the inputs.

First assume that $\alpha<1/4$. Clearly $k$ is an upper bound on
$k'=bound^{(1)}(f)$ in this case.
We have to show that also $k=O(k')$.

Let $\mu$ be an $\alpha$-balanced distribution on the inputs. We
can balance the weights of 1-inputs and 0-inputs by multiplying the
weights of the 0-inputs by some $p_0$ and the weights of the 1-inputs
by some $p_1$ so that a strictly balanced distribution $\mu'$ is
obtained. We take a rectangle $R$ of error $\epsilon$ and size $1/2^{k'}$
according to $\mu'$. Assume that
$\alpha \le\mu(f^{-1}(1))\le 1/2$. Then $p_1>1$. Consequently the size of $R$
is slightly smaller according to $\mu$ than to $\mu'$, and the error
is possibly slightly smaller, too. In the other case $\alpha
\le\mu(f^{-1}(0))\le 1/2$ the opposite
occurs, namely the rectangle is possibly slightly larger, and the
error is larger. But in any case the size and the error are changed by constant
factors $\Theta(\alpha)$ resp.~$\Theta(1/\alpha)$ only. If the error
of $R$ is too
large we can use the previous
lemma to reduce the error probability while decreasing the size.

The case $1/4\le\alpha\le 1/2$ is handled similarly.
\qed

\section{The rectangle bound versus Arthur Merlin communication
 complexity}

We now prove the relations between Arthur Merlin communication and the
lower bound method based on rectangle size.

{\bf Proof of Theorem~\ref{the:AMMABD}, part 1.}
We are given a Merlin Arthur protocol with complexity $c$ and error
$\epsilon\le 1/4$ for the function $f$. We show that under this condition we can
find a large $1-\epsilon$-correct 1-rectangle.
Recall that the complexity of the Merlin Arthur protocol
includes the ``communication'' done by Merlin (who is guessing
nondeterministically) plus the communication by the
players Alice and Bob. Let $c_M$ be the length of the longest guess
given by Merlin over
all inputs. We will call such a guess string a {\it proof}. Let $c_P$
denote the length of the longest communication between Alice and Bob occuring
during any run of the protocol. Clearly $c_M,c_P\le c$.

It is possible to reduce the error probability of the protocol to
$1/2^{2c}$ by repeating the probabilistic part of the protocol
$O(c)$ times independently and taking the majority output, for any
fixed proof of Merlin. Let
$P(x,y,z)$ denote the (random) output of the protocol for inputs $x,y$
and Merlin's proof $z$.
The acceptance properties of the protocol are then:

If $f(x,y)=1$ then there is a proof $z$ so that $P(x,y,z)$ accepts
with probability $1-1/2^{2c}$.

If $f(x,y)=0$ then for all proofs $z$, $P(x,y,z)$ accepts
with probability at most $1/2^{2c}$.

Note that the communication among the players in the new protocol is
bounded by $k=O(c\cdot c_P)=O(c^2)$.

There are at most $2^{c_M}$ different proofs. If we fix such a proof $z$
there is a set $s_z$ of 1-inputs that is accepted on this proof, i.e., for
which the protocol accepts with high probability on this proof. In
this way the set of 1-inputs is covered by $2^{c_M}$ subsets
$s_1,\ldots,s_{2^{c_M}}$.

Let $\mu$ be any balanced distribution over the inputs. For each such distribution we
can find at least one proof $z$ so that $\mu(s_z)\ge 1/2^{c_M}$.
We fix such a proof. This turns the
Merlin Arthur protocol into a randomized protocol so that a subset
$s_z$ of 1-inputs of weight $1/2^{c_M}$ is accepted with probability
$1-1/2^{2c}$ each and no
0-input is accepted with probability larger than $1/2^{2c}$. The other 1-inputs
are accepted with uncertain probability.

We restrict $\mu$ to the
inputs in $s_z\cup f^{-1}(0)$, by setting the weight of all other
inputs to 0 and normalizing to a distribution $\mu'$. Clearly the
error of the protocol under $\mu'$ is at most $1/2^{2c}$. Furthermore
note that either \begin{equation}\label{eq:balMA}
\mu'(x,y)=0 \mbox{ or }\mu'(x,y)=\Theta(\mu(x,y)),\end{equation} since
$\mu$ is balanced.

Given such a randomized protocol
we may also fix its random choices and get a deterministic protocol (like
in the easy direction of the Yao-principle).
This yields a deterministic protocol which on expectation
has error $1/2^{2c}$ (under $\mu'$).
Consequently there exists a deterministic protocol with
error $1/2^{2c}$ under $\mu'$ and communication $k=O(c^2)$.

A deterministic protocol with communication $k$ easily leads to a set $R$
of $P=2^{O(k)}$ pairwise disjoint rectangles labeled with the protocol
output that
partition the communication matrix.
We show that there exists a large 1-rectangle with small error.

Assume that all rectangles which are larger than $1/(2^{c+1}\cdot 2P)$ have error
larger than $1/2^{c}$. Then the success probability of the protocol on $\mu'$
is upper bounded as follows. The small rectangles contribute at most
$P\cdot 1/(2^{c+1}\cdot 2P)\cdot1\le 1/2^{c+2}$, the large rectangles all have
success at most $1-1/2^{c}$ and so the overall success probability is at most
$1-1/2^{c}+ 1/2^{c+2}$, too small in comparison to the maximum
error $1/2^{2c}$. Hence there is a 1-rectangle of size
$1/(2^{c+1}\cdot 2P)\ge 2^{-\Omega(c^2)}$ with error at most
$1/2^{c}\le\epsilon$ according to $\mu'$. If we switch from $\mu'$ to
$\mu$, then the size
of a rectangle cannot decrease (compared to
$\mu$) by more than a
constant factor due to (\ref{eq:balMA}). It is also easy to see that
the error of the rectangle
cannot increase when switching from $\mu'$ to $\mu$.
Consequently $bound^{(1)}_\epsilon(f)\le O(c^2)$.
\qed

Now we relate $bound^{(1)}(f)$ to $AM(f)$.

{\bf Proof of Theorem~\ref{the:AMMABD}, part 2.}
Assume that $bound^{(1)}_{\epsilon}(f)=c$. Then
$bound^{(1)}_{\epsilon^4/8}\le O(c)$. 
For {\it all balanced} distributions $\mu$ there is a rectangle $R_\mu$ with
error at most $\epsilon^4/8$ and size at least
$s\ge 1/2^{O(c)}$. Also recall that $AM_\epsilon(f)=\max_\mu
N_\epsilon^\mu(f)$, where $\mu$
runs over {\it all} distribution on the inputs, due to Lemma~\ref{lem:AMdistr}.
We use a greedy algorithm to construct a cover of the 1-inputs to $f$ with
error $\epsilon$ containing at most
$2^{O(c)}\cdot(1/\epsilon^2)\cdot\log(1/\epsilon)$ rectangles for any
$\mu$.

So let $\mu$ be some distribution on $\dom$. 
We distinguish three cases. First consider the case that
$\mu(f^{-1}(1))\le \epsilon^2$. In this case clearly
$N_{\epsilon^2}^\mu(f)=0$ by a protocol that never accepts.

Next consider the case that
$\mu(f^{-1}(0))\le \epsilon^2$. In this case $N_{\epsilon^2}^\mu(f)=0$ by a
protocol that always accepts.

Now consider the case that $\mu(f^{-1}(1))\ge\epsilon^2$ and
$\mu(f^{-1}(0))\ge\epsilon^2$. Then
we can still find a good cover
as follows. We first show that for each such distribution a relatively
large rectangle with small error exists. Then we use a greedy approach
to find a cover.

First we show how to find good rectangles. We (strictly) balance the distribution
by multiplying the weights of
1-inputs by some value $p_1$ and multiplying the weights of
0-inputs by some value $p_0$. Clearly
\[2(1-\epsilon^2)\ge \frac{1}{p_1},\frac{1}{p_0}\ge 2\epsilon^2.\]
For the resulting strictly balanced distribution
$\mu'$ there is a 1-rectangle $R_{\mu'}$ of size $s$ having error
$\epsilon^4/8$ at most. Then the $\mu$-weight of 1-inputs in $R_{\mu'}$ is at least
\[s\cdot(1-\epsilon^4/8)\cdot (1/p_1)\ge
s\cdot(1-\epsilon^4/8)\cdot2\epsilon^2\ge s\epsilon^2.\]
Furthermore the $\mu$-weight
of 0-inputs in $R_{\mu'}$ is at most \[s\cdot(\epsilon^4/8)\cdot (1/p_0)\le
s\cdot\epsilon^4 \cdot (1-\epsilon^2)/4\le s\epsilon^4/4.\]

We next construct for $\mu$ a cover using a greedy
approach.

\begin{enumerate}
\item Let $\mu_0=\mu$.
\item For $\mu_i$ find a rectangle $R_i=R_{\mu_i}$ that contains
 1-inputs of weight at least $s\epsilon^2$ and 0-inputs of weight at
 most $s\epsilon^4/4$.
\item Put $R_i$ into the
cover.
\item Remove the weight from all 1-inputs in $R_i$ and uniformly
increase the weights of the remaining 1-inputs by some appropriate
factor $q(i)$. [Note that this does not affect the balance of the
distribution.] Let $\mu_{i+1}$ denote the resulting distribution. 
\item Stop,
when the set of remaining 1-inputs not covered so far has weight
$\le\epsilon/2$ according to $\mu$.
\item Otherwise continue with 2.~and set $i:=i+1$.
\end{enumerate}

Clearly the algorithm finds a set of rectangles so that all but a set
of weight $\epsilon/2$ of the 1-inputs is covered. In the worst
case the weight of any 1-input is increased by a factor of $q(1)\cdots
q(i)\le(3/4)/(\epsilon/2)=
(3/2)/\epsilon$ during the course of the algorithm. Hence the weight of
1-inputs in $R_i$ according to $\mu$ is at least
$s\epsilon^2\cdot\epsilon\cdot (2/3)$, while the weight of 0-inputs is
at most $s\epsilon^4/4$.
The error of $R_i$ is thus at most $(3/8)\epsilon$. Since this holds for all
rectangles, the weight of 0-inputs in the cover is at most a fraction
of $(3/8)\epsilon$ of the weight of all 1-inputs covered, which is at
most $(3/4)\cdot (3/8)\epsilon<\epsilon/2$. The weight of 1-inputs
not covered is $\epsilon/2$.
So the obtained cover has error $\epsilon$.

Now we have to analyze the size of the obtained cover. Each step
covers at least a $s\epsilon^2$ fraction of the remaining 1-inputs. Hence the proportion
of not yet covered 1-inputs according to $\mu$ after $k$ steps is
\[w_k\le(1-s\epsilon^2)^k.\]
The algorithm stops if this is smaller than $O(\epsilon)$, hence
$k\le O(1/(s\epsilon^2)\cdot\log(1/\epsilon))$, and
$N_{\epsilon}^\mu(f)\le O(c+\log(1/\epsilon))$.

So indeed for all $\mu$ we have $N_\epsilon^\mu(f)\le
O(bound^{(1)}_\epsilon(f)+\log(1/\epsilon))$, and hence
$AM_\epsilon\le O(bound^{(1)}_\epsilon(f)+\log(1/\epsilon))$.
\qed

{\bf Proof of Theorem~\ref{the:distribconj}.} 
We first define the function that has both small nondeterministic and
co-nondeterministic complexity with 0 error under some distribution, but large
deterministic complexity for some constant error under the same distribution.

Let $WHICH((x_1,x_2),(y_1,y_2))=$ \[\left\{\begin{array}{ll}
1 \mbox{ if } \neg DISJ(x_1,y_1)=1 \mbox{ and } \neg DISJ(x_2,y_2)=0\\
0 \mbox{ if } \neg DISJ(x_1,y_1)=0 \mbox{ and } \neg DISJ(x_2,y_2)=1\\
0 \mbox{ otherwise.}\end{array}\right.\]

We employ the following more specific and optimized version of
Fact~\ref{fac:Razb}, which follows from some fine-tuning of the result
in \cite{R92}.
\begin{fact}\label{fac:Razb2}
Let $\nu_a$ be the distribution on $\dom$
which is uniform on $\{(x,y):|x|=|y|=n/4, |x\cap
y|=0\}$ (and 0 elsewhere), and let $\nu_r$ be the distribution on $\dom$
which is uniform on $\{(x,y):|x|=|y|=n/4, |x\cap
y|=1\}$ (and 0 elsewhere). Let $\nu$ be the distribution on $\dom$, which is defined by
$\nu(x,y)=(3/4)\cdot\nu_a(x,y)+(1/4)\cdot \nu_r(x,y)$.

Then for any constant $\delta>0$ there is a constant $\beta(\delta)>0$,
so that any rectangle $R$ either 
has size $2^{-\beta(\delta) n}$,
or
$\nu(x,y:x\cap y\neq\emptyset\,|\,R)\ge 1/4-\delta$, i.e.,
\[\nu(\{(x,y):x\cap y\neq\emptyset\}\cap R)\ge (1/4-\delta) \cdot\nu(R)-2^{-\beta(\delta) n}.\]
\end{fact}

Now to the definition of the distribution on the
inputs. In the distribution $\nu\times\nu$ two instances $(x_1,y_1)$
and $(x_2,y_2)$ are chosen independently from $\nu$.

For the hard distribution $\mu$ on inputs we pick inputs as in
$\nu\times\nu$, but inputs with
$DISJ(x_1,y_1)= DISJ(x_2,y_2)$ are removed, so
that the function value on the two instances differs with probability
1 ($\mu$ is normalized to a distribution after this removal).
On $\mu$ the task of a protocol is to determine on which of the two
set pairs $\neg DISJ$ is true. Note that either
$\mu(x_1,x_2,y_1,y_2)=\Theta(\nu\times\nu(x_1,x_2,y_1,y_2))=\Theta(\nu(x_1,y_1)\cdot\nu(x_2,y_2))$
or $\mu(x_1,x_2,y_1,y_2)=0$, for all inputs $(x_1,x_2,y_1,y_2)$ from $\{0,1\}^{4n}$,
since $Prob_{\nu}(DISJ(x_1,y_1)=1)=3/4.$
Also note that $\mu$ is strictly balanced.

There is a simple nondeterministic protocol for $WHICH$ making no error under the
distribution $\mu$. One can simply use a protocol for
$\neg DISJ$ on the first instance. This covers all 1-inputs of $WHICH$,
but accepts no 0-input with weight larger than 0. Analogously we can
find a protocol for $\neg WHICH$ under $\mu$. So
$N_0^\mu(WHICH),N_0^\mu(WHICH)\le\log n+1$.

Now we turn to the complexity of a deterministic protocol with
error. Such a protocol with communication $c$ immediately yields a set
of $P=2^{O(c)}$ pairwise {\it disjoint} rectangles labeled with values 0,1, so that
a $1-\epsilon$ fraction of all inputs according to $\mu$ are in correctly labeled
rectangles. Call the 1-rectangles $R_1,\ldots,R_P$, the 0-rectangles $S_1,\ldots,S_P$.

We show that such a partition can only exist if
$c=\Omega(n)$. So for the sake of contradiction assume that $c\le
\gamma n$ for some arbitrarily small constant $\gamma$ we can choose later.

Note that the difficulty for a deterministic protocol is that the
corresponding cover consists of disjoint rectangles, even on the
inputs with $\mu(x_1,x_2,y_1,y_2)=0$. In case the reader would prefer to loosen
this restriction and require a protocol to be deterministic only on
those inputs with $\mu(x_1,x_2,y_1,y_2)>0$ we can still give those inputs some
small probability, so that the above nondeterministic protocols would have
small error, while the following lower bound on
deterministic protocols would be unchanged.

We use the following notation. An input
$(x_1,x_2,y_1,y_2)$ is in {\it quadrant} A, if $\neg DISJ(x_1,y_1)=0$
and $\neg DISJ(x_2,y_2)=0$, in quadrant B, if $\neg DISJ(x_1,y_1)=1$
and $\neg DISJ(x_2,y_2)=0$, in quadrant C, if $\neg DISJ(x_1,y_1)=0$
and $\neg DISJ(x_2,y_2)=1$, and in quadrant D, if $\neg DISJ(x_1,y_1)=1$
and $\neg DISJ(x_2,y_2)=1$. Note that inputs in quadrants A and D have
probability 0 under $\mu$. Under $\nu\times \nu$ quadrants B and C
have weight $3/16$, quadrant A has weight $9/16$ and quadrant D has
weight $1/16$.

Let $R_i$ be a 1-rectangle and $(x_1,x_2,y_1,y_2)\in R_i$ with $\neg
DISJ(x_1,y_1)=1$. Then the set
$R_i(x_1,y_1)=\{x_2,y_2: x_1,x_2,y_1,y_2\in R_i\}$ is a rectangle in
$\{0,1\}^n\times\{0,1\}^n$.
Let \[\mu(R_i(x_1,y_1)\,|\,x_1,y_1)=\frac{\mu(\{(x_1,y_1)\}\times
 R_i(x_1,y_1))}{\mu(\{(x_1,y_1)\}\times\dom)}\]
denote the weight of $R_i(x_1,y_1)$ relative to the inputs with fixed
$x_1,y_1$. Note that
\[\mu(\{(x_1,y_1)\}\times\dom)=\Theta(\nu(x_1,y_1)).\]
Also
\[\mu(\{(x_1,y_1)\}\times R_i(x_1,y_1))\le O(\nu(x_1,y_1)\cdot\nu(R_i(x_1,y_1)))
,\]
since all inputs $x_1,x_2,y_1,y_2$ with $\neg DISJ(x_2,y_2)=1$ have
weight 0 in $\mu$ and all $x_1,x_2,y_1,y_2$ with $\neg DISJ(x_2,y_2)=0$ have
weight $\mu(x_1,x_2,y_1,y_2)=\Theta(\nu(x_1,y_1)\cdot\nu(x_2,y_2))$.
Then
\begin{equation}\label{eq:rectw}
\nu(R_i(x_1,y_1))=\Omega(\mu(R_i(x_1,y_1)\,|\,x_1,y_1)).\end{equation}

We will show that each large $R_i(x_1,y_1)$ must contain many inputs $x_2,y_2$ with
$\neg DISJ(x_2,y_2)=1$. While this does not
create any error in the rectangle $R_i$, this shows that $R_i$ occupies a significant
portion of quadrant D.
If this is true for many rectangles $R_i$ then a situation is
reached in which the
majority of quadrant D is occupied. Since a symmetric argument applies
to the 0-rectangles we are lead into a contradiction, since the
rectangles are not allowed to intersect nontrivially.

We set $\epsilon=\delta/2=1/17$ fixing the protocol's error and the
constant from Fact~\ref{fac:Razb2}, and 
choose $\gamma<\beta(\delta)/2$. Let $x_1,y_1$ be an input with $\neg DISJ(x_1,y_1)=1$.
If $\mu(R_i(x_1,y_1)|x_1,y_1)\ge \Omega(2^{-\gamma n})$, then with (\ref{eq:rectw})
$\nu(R_i(x_1,y_1))\ge 2^{-\beta(\delta) n}$, and hence
the fraction of $x_2,y_2\in
R_i(x_1,y_1)$ with $\neg DISJ(x_2,y_2)=1$ is at least $1/4-\delta$ according to
$\nu$.
In other words, the proportion of 0-inputs to 1-inputs of $\neg
DISJ(x_2,y_2)$ in $R_i(x_1,y_1)$ is $3/4+\delta$ to $1/4-\delta$.

Hence $\{(x_1,y_1)\}\times R_i(x_1,y_1)$ occupies
at least a $(1/4-\delta)/(3/4+\delta)\ge 1/3-2\delta$ fraction of
the weight it covers in quadrant B (according to $\nu\times \nu$)
also on the inputs in quadrant D (according to
$\nu\times\nu$). Recall that on $\nu\times\nu$ quadrant $D$ has weight $1/16$ and
quadrant $B$ has weight $3/16$.

We next show that a fraction of $1-2\epsilon$
of all 1-inputs of $WHICH$ (resp.~of quadrant B) is covered by
$R_i(x_1,y_1)$ with $\mu(R_i(x_1,y_1)|x_1,y_1)\ge\Omega(\epsilon\cdot
2^{-\gamma n})$ and hence $\nu(R_i(x_1,y_1))\ge 2^{-\beta(\delta)
 n}$ (using (\ref{eq:rectw})). Then in quadrant D weight at least
$(1-2\epsilon)\cdot(1/3-2\delta)\cdot 3/16>1/16-\delta/2>(1/2)\cdot (1/16)$ is occupied by the
rectangles $R_i$ (according to $\nu\times \nu$). Since the same
holds for the rectangles $S_i$ we get
at least one input in quadrant D that is covered twice, a contradiction
to the requirements on those rectangles. Therefore
$\gamma\ge\beta(\delta)/2=\Theta(1)$ and hence $c=\Omega(n)$.

The weight of 1-inputs (according to $\mu$) that are in rectangles
$R_i(x_1,y_1)$ with $\mu(R_i(x_i,y_i)\,|\,x_1,y_1)\ge
\epsilon/const\cdot 2^{-\gamma n}$ (and $\neg DISJ(x_1,y_1)=1$) is at
least $1-2\epsilon$,
since at most an $\epsilon$ fraction of 1-inputs is not covered and
the fraction of 1-inputs covered by smaller rectangles can be
bounded as follows. Each small rectangle $R_i(x_1,y_1)$ can cover
1-inputs of weight at most
$\epsilon/const\cdot 2^{-\gamma n}\cdot
\mu(\{(x_1,y_1)\}\times\{0,1\}^n\times\{0,1\}^n)$, and there are at
most $2^{\gamma n}\cdot S$ different $R_i(x_1,y_1)$ with $\neg DISJ(x_1,y_1)=1$, where
$S=|\{(x,y):|x|=|y|=n/4, |x\cap y|=1\}|$. Note that
$\mu(\{(x_1,y_1)\}\times\{0,1\}^n\times\{0,1\}^n)=\Theta(1/S)$.
So the 1-inputs covered by small rectangles have weight at most
\[\epsilon/const\cdot 2^{-\gamma n}\cdot
 \mu(\{(x_1,y_1)\}\times\{0,1\}^n\times\{0,1\}^n)\cdot 2^{\gamma n}
 \cdot S\le
 \epsilon.\,\Box\]

\section{Uniform threshold covers and the rectangle bound}

{\bf Proof of Theorem~\ref{the:BDUT}, part 1.}
Assume that $UT^{(1)}_{s,2s}(f)\le k$ for some $s$. Then following
Remark~\ref{rem:boostUT},
$UT^{(1)}_{\epsilon s',s'}(f)\le O(k)$ for an
arbitrarily small constant $\epsilon$ and some $s'$.

Given a one-sided bounded error uniform threshold cover with $P\le2^{O(k)}$
1-rectangles $S=\{R_1,\ldots,R_P\}$ let $h(x,y)$ denote the number
of rectangles $R_i$ the input $x,y$ is included in.

We know that for each $x,y$ with $f(x,y)=1$ there are at least $s'$
1-rectangles it is included in, so $h(x,y)\ge s'$. Each $x,y$ with
$f(x,y)=0$ is in at most $\epsilon s'$ 1-rectangles, hence $h(x,y)\le
\epsilon s'$.

Let $\mu$ be any balanced distribution on the inputs. We define a
probability distribution $\nu$ on the 1-rectangles in $S$ as follows. Each rectangle
$R\in S$ receives the
weight $\sum_{x,y\in R} \mu(x,y)$. We then normalize these
weights to a distribution on 1-rectangles in $S$.
The probability of
some rectangle $S$ is then \[\sum_{x,y\in R}
\frac{\mu(x,y)}{\sum_{R'}\sum_{x',y'\in R'}\mu(x',y')}=\sum_{x,y\in R}
\frac{\mu(x,y)}{\sum_{x',y'\in\dom}\mu(x',y')\cdot h(x',y')}.\]

If we first pick a rectangle according to $\nu$ and then on that
rectangle an input (according to $\mu$ restricted to $R$), we get some
input $x,y$ with probability
\[\frac{\mu(x,y)\cdot h(x,y)}{\sum_{x',y'\in\dom}\mu(x',y')\cdot h(x',y')}.\]

So the weight of $x,y$ in this experiment
is proportional to $\mu(x,y)\cdot h(x,y)$. Hence the probability of
picking a 0-input in this way is at most
\begin{equation}\label{eq:err}
\frac{\mu(f^{-1}(0))\cdot \epsilon s'}
{\mu(f^{-1}(1))\cdot s'}\le\frac{3/4\cdot \epsilon s'}{1/4\cdot s'}\le
3\epsilon.\end{equation}

Assume that all rectangles that are larger than $\epsilon^2/P$ according to
$\mu$ have error larger than $4\epsilon$. Then, if we first pick a
rectangle $R$
and then an input $x,y\in R$ the probability that $f(x,y)=1$ can be
bounded as follows. The small rectangles
contribute at most $P\cdot (\epsilon^2/P)\cdot 1\le\epsilon^2$ to this probability.
All larger rectangles have error $4\epsilon$ at least, and hence when
picking one of them the probability of getting a 1-input is at most $1-4\epsilon$,
so the overall probability of getting a 1-input is at most
$1-4\epsilon+\epsilon^2$, a contradiction to~(\ref{eq:err}).

Hence there exists a 1-rectangle of size at least $\epsilon^2/P=\Omega(1/P)$ having error at
most $O(\epsilon)$ according to $\mu$.

When given a bounded error uniform threshold cover we can do the same construction for
the 0-inputs, and hence $UT_{s,2s}(f)=k$ allows us to find both
a 1-rectangle and a 0-rectangle with the desired properties for any
balanced $\mu$.
\qed

{\bf Proof of Theorem~\ref{the:BDUT}, part 2.}

Assume $bound^{(1)}_{1/4}(f)=k$ for some $f:\fdom$. Then $bound^{(1)}_{1/n^5}(f)\le
O(k\log n)$ using Lemma~\ref{lem:err-red}. In other words, for each balanced distribution $\mu$ on
$\dom$ there exists a $1-1/n^5$-correct 1-rectangle of size at least
$s=2^{-O(k\log n)}$.

We show how to construct a one-sided uniform
threshold cover with parameters $n,n^2$.
The cover is produced by an algorithm. Let $\mu_1$ be the distribution
which is uniform on the 1-inputs of $f$ with probability 1/2 and
uniform on the 0-inputs of $f$ with probability 1/2.

\begin{enumerate}
\item Set $l=1$, $Cov_0=\emptyset$.
\item Find a $1-1/n^5$-correct rectangle $R_l$ of size at least $s$
 according to $\mu_l$ and let
 $Cov_l=Cov_{l-1}\cup \{R_l\}$.
\item Let $I_0(l)$ denote the set of 0-inputs in $R_l$, let $I_1(l)$
 denote the set of 1-inputs in $R_l$.
\item Construct $\mu_{l+1}$ be as follows: 
\begin{itemize}
\item the weight of all inputs in
 $I_1(l)$ is reduced by a factor of $1-1/n^4$. The obtained ``free''
 weight is used to increase the weights 
 of all inputs in $I_0(l)$ by a fixed factor.
\item Any input in $I_1(l)$ that is covered more than $n^2$ times
 receives weight 0. Its weight is used to increase the weight of all
 1-inputs by a fixed factor.
\end{itemize}
\item STOP if all 1-inputs are covered at least $n^2$ times.
\end{enumerate}

Note that each 0-input that is covered in some iteration $l$ at least doubles its
weight at that point. Namely, since a rectangle $R_l$ has error
$1/n^5$, weight $\mu_l(R_l)\cdot(1-1/n^5)\cdot 1/n^4$ is
distributed to inputs in $I_0(l)$ of
weight $\mu_l(R_l)\cdot 1/n^5$, more than doubling the weight of
each such input. Since no 0-input has weight more than $1$ or less
than $1/2^{2n}$ this implies that no 0-input is ever covered more than $2n$ times.

We have to show that the distributions $\mu_l$ are all
balanced for step 2.~to work. The second part of step 4.~does not
change the balancedness of the distribution. In the first part of step 4.~some
weight is shifted from 1-inputs to 0-inputs. The inputs in $I_1(l)$
are reduced in weight by a factor of $(1-1/n^4)$. But as soon as an
input is covered $n^2$ times this reductions stops.

Let us
assume for the moment the
following lemma justifying step~2., whose proof will be provided at the
end of this section.

\begin{lemma}\label{lem:bal2}
$1/2\ge\mu_l(f^{-1}(1))\ge 1/2-O(1/n)$ for all $l$.
\end{lemma}

The lemma clearly implies that the distributions $\mu_l$ are all
balanced, hence step 2.~is applied correctly. We use
Lemma~\ref{lem:bal2} only to ensure that the obtained cover is good on the
0-inputs, but not to ensure that it is good on the 1-inputs.
To analyze the number of iterations of the algorithm we
consider the following modification of step 2.

\begin{enumerate}
\item[2.'] First strictly balance $\mu_l$ by
uniformly increasing the weights of 1-inputs and decreasing the weights
of 0-inputs by fixed factors. Then pick a size $s$ rectangle $R_l$ with
error $1/n^5$ according to that distribution.
\end{enumerate}

The size of $R_l$ on $\mu_l$ is at
least $s\cdot d_l$, when
$d_l=\mu_l(f^{-1}(1))/(1/2)$ denotes the distortion of the balance
of $\mu_l$ compared to $\mu_1$. Note that $\mu_l(f^{-1}(1))\le 1/2$,
so $d_l\le 1$.

We will show that the modified algorithm terminates and produces a
size $O(n^3/s)$ cover. This immediately implies that each 1-input is
covered at least $n^2$ times.
Then this is also true for
the original algorithm: The original algorithm uses larger rectangles
in step 2.~and hence terminates faster. 
Furthermore we will show that Lemma~\ref{lem:bal2} holds for both the modified and the
original algorithm. Note that, however, only the original algorithm
guarantees that the cover is good on the 0-inputs.

Let $S_l$ denote the set of 1-inputs not covered $n^2$
times before iteration $l$, and let $N_l=|S_l|$. Since the weight of each input in
$S_l$ is reduced in the first part of step 4.~by a factor of $(1-1/n^4)$ each time
it is covered, this decreases the weight of such an input by
$(1-1/n^4)^{n^2-1}\le (1-O(1/n^2))$. In the second part of step 4.~the
weights of all 1-inputs are increased by some fixed factor. Furthermore
$\mu_1(x,y)=1/(2N_1)$ for all 1-inputs $x,y$. Hence for all $x,y\in
S_l$ and $x',y'\in S_l$:
\[(1-O(1/n^2))\cdot \mu_l(x',y')\le\mu_l(x,y)\le (1+O(1/n^2))\cdot\mu_l(x',y').\]
The average weight of an
input in $S_l$ is $(1/(2N_l))\cdot d_l$.
Hence the for {\it all} $x,y\in S_l$
\begin{equation}\label{eq:weight} (1/(2N_l))
 d_l(1-O(1/n^2))\le\mu_l(x,y)\le (1/(2N_l))
 d_l(1+O(1/n^2)).\end{equation}

Let $h_l(x,y)$ denote the number of times input $x,y$ is covered by
$Cov_{l-1}$. There are $N_1$ 1-inputs to $f$. Then let
\[h(l)=\sum_{(x,y)\in f^{-1}(1)} 1/N_1\cdot pos(n^2-h_l(x,y)),\] where
$pos(x)=x$ if $x\ge 0$ and $pos(x)=0$ otherwise.
$h(l)$ denotes the average number of times 1-inputs still have to be
covered. Clearly $n^2\ge
h(l)\ge 0$, and if $h(l)>0$, then $h(l)\ge 1/N_1$. 

In each step 1-inputs of weight $s(1-1/n^5)d_l$ according to $\mu_l$ are
covered. Let $C_l$ denote the set of 1-inputs $x,y\in R_l$ with $h_{l}(x,y)<n^2$.
Due to~(\ref{eq:weight}): $\mu_1(C_l)\ge sd_l(1-1/n^5)\cdot (1-O(1/n^2))\cdot(1/d_l) 
N_l/N_1\ge s/2\cdot N_l/N_1$.
Including $R_l$ in the cover reduces $h(l)$ by at least $sN_l/N_1$ hence.
Then
\begin{eqnarray*}
h(l+1)&\le& h(l)-sN_l/N_1\\
&=&\sum_{(x,y)\in f^{-1}(1)} 1/N_1\cdot pos(n^2-h_l(x,y))
-(N_l/N_1)\cdot s\\
&=&\sum_{(x,y)\in S_l} 1/N_1\cdot (n^2-h_l(x,y)
-s)\\
&\le &\sum_{(x,y)\in S_l} 1/N_1\cdot
(n^2-h_l(x,y)) \cdot(1- s/n^2)\\
&\le &h(l)\cdot(1- s/n^2).\end{eqnarray*}
For some $l=O(n^3/s)$ iterations
$h(l)=0$. Hence the constructed cover contains no more than $O(n^3/s)$
rectangles. Since the algorithm with the original step~2.~terminates
at least as fast we have
$UT^{(1)}_{n,n^2}(f)\le O(k\log n)$.

Given that $UT_{s,t}(f)=\max\{UT^{(1)}_{s,t}(f),UT^{(1)}_{s,t}(\neg f)\}$ we can
simply do the same construction for the 0-inputs and get the desired
result for $bound(f)$.
\qed

{\bf Proof of Lemma~\ref{lem:bal2}.} First let us look at the
algorithm with the modified step~2'.
Let $s_ld_l$ denote the weight of $R_l$ in $\mu_l$.
As argued before, the algorithm stops as soon as $\prod (1-s_l/n^2)< 1/(n^2\cdot N_1)$.
So there is a sequence $s_1,\ldots, s_k$ so that
$\prod_{l=1}^k (1-s_l/n^2)<1/(n^2\cdot N_1)$ and all $s_l\ge s$, and $k$ is minimal with
this property. The weight transferred to the 0-inputs is then at
most $\sum_{l=1}^{k-1} s_ld_l/n^4\le\sum_{l=1}^{k-1} s_l/n^4$, since no weight is transferred from
$R_k$. We may hence adjust $s_k$ so that $\prod_{l=1}^k
(1-s_l/n^2)=1/(n^2\cdot N_1)$.

For each $k$ it is true that
$\sum_{l=1}^k s_l$ is maximized for $s_1=\cdots=s_k=:\bar{s}$ because: let
$s_l'=(1-s_l/n^2)$. Then $\prod s_l'=1/(n^2N_1)$ and we want to maximize
$\sum (1-s_l'n^2)=k-n^2\sum s_l$ or equivalently minimize $\sum
s_l$. This is achieved when $s_1=\cdots=s_k$.

Then $k=O(n^3/\bar{s})$ and the transferred weight is at most $k\cdot
\bar{s}/n^4=O(1/n)$. Consequently the
same holds if $k$ is arbitrary. Note that this implies $d_l\ge
1-O(1/n)$.

In case the original step 2.~of the algorithm is applied
$\mu_l(R_l)=s_l\ge s$, and potentially
a larger weight is transferred to the 0-inputs. But this also makes
the algorithm terminate quicker. 

The algorithm stops at least when $\prod (1- s_l/(d_ln^2))< 1/(n^2\cdot N_1)$.
We may substitute $s_l''=s_l/d_l$ and are left with the problem of
finding the
maximum of $\sum s_l''d_l$ under the constraint that
$\prod (1-s_l''/n^2)=1/(n^2\cdot N_1)$ and $s_l''=s_l/d_l\ge s/d_l\ge s$. This is
the problem we have just analyzed.
\qed

\section{Comparing the power of different threshold covers}

First let us show that one-sided bounded error uniform threshold covers for some
function $f$ can easily be converted into
approximate majority covers. The same also holds for the complement of $f$.

{\bf Proof of Theorem~\ref{the:APPUT}.}
Assume that $UT^{(1)}_{t,9t}(f)=k$. Then there exist $2^k$ rectangles so
that each 1-input is in at least $9t$ rectangles and each 0-input is
in at most $t$ rectangles. Now label all the rectangles as
1-rectangles and add $3t$ times the 0-labeled rectangle covering all
inputs. This is clearly an approximate majority cover, hence $APP(f)\le
UT^{(1)}_{t,9t}(f)\le O( UT^{(1)}_{t,2t}(f))$.
\qed

Now we relate $APP(f)$ to a version of the rectangle size bound.

{\bf Proof of Theorem~\ref{the:APPrect}.}
Assume that $APP(f)=k$, then we can find $2^k$ labeled rectangles
making up an approximate majority cover for $f$. We first have to
show that in this case for
each balanced distribution $\mu$ there exists a $3/4$-correct
rectangle of size $1/2^{O(k)}$ at least. The proof is analogous to the
proof of Theorem~\ref{the:BDUT}.1.a, but this time we are
guaranteed to find a large rectangle
with small error, not a large 1-rectangle with small error. To adapt
the proof one has to replace the uniform threshold values $s', \epsilon
s'$ by the expected correct height $(1-\epsilon)\cdot E[h(x,y)]$ and
incorrect height $\epsilon\cdot E[h(x,y)]$.

Now we show the opposite direction, namely, given that for each
balanced distribution $\mu$ we
can find a $3/4$-correct rectangle of size $1/2^k$, then
we can construct an approximate majority cover.

First notice that in fact we can find a rectangle of size $1/2^k$ and
error at most $1/4$ for {\it all} distributions on the inputs, since
on unbalanced distributions we may simply take $\dom$ as a rectangle
with error $1/4$ when choosing the appropriate label for that rectangle.

To construct the approximate majority cover we first consider the fact that $\min_\mu \max_v
size(\mu,\epsilon,f,v)\ge 1/2^k$ in a somewhat different
light. Let
\[para(\mu,R)=\epsilon/(3\mu(R))+err(R,\mu,v
(R))\cdot 2^k.\]
This parameter controls the quality of a rectangle.
Yao's application of the minimax-principle to randomized
algorithms (see \cite{KN97}) provides us with the following statement.

\begin{lemma}\label{lem:Yao}
The following two statements are equivalent for all $f$.
\begin{enumerate}
\item For all distributions $\mu$ there is a rectangle $R_\mu$ with parameter
 $\alpha$.
\item There is a probability distribution $D$ on rectangles so that for
 all distributions $\mu$ on inputs the expected parameter of a rectangle
 is $\alpha$.
\end{enumerate}
\end{lemma}

The latter could be named a ``randomized rectangle'' because it
resembles a randomized algorithm. We know that for all distributions
$\mu$ on $\dom$ there is a rectangle $R_\mu$ with size $1/2^k$ and error $\epsilon=1/4$, hence
$para(\mu,R_\mu)\le (1/4)/(3/2^k)+(1/4)\cdot2^k=(1/3)\cdot 2^k$.
The randomized rectangle then offers a distribution on
rectangles with expected parameter $(1/3)\cdot 2^k$ at most. Hence the
expected rectangle size satisfies $(1/4)/(3 E[\mu(R)])\le 
(1/3)\cdot 2^k\Rightarrow E[\mu(R)]\ge 1/2^{k+2}$. The expected
error satisfies $E[err(R,\mu,v(R))]\cdot 2^k\le (1/3)2^k\Rightarrow E[err(R,\mu,v(R))]\le 1/3$.

A randomized rectangle immediately gives us an approximate majority cover for
$f$, though not of the desired
size. To see this note that if we consider a distribution $\mu_{x,y}$
concentrated on some fixed input $x,y$,
then $Prob_D(v(R)\neq f(x,y)|x,y\in R)= E[err(R,\mu_{x,y},v(R))]\le 1/3$.

From the direct application of the Yao-principle we do not get a bound
on the number of rectangles with nonvanishing probabilities used in $D$.
We use the following discretization.

\begin{lemma}
Assume there is a randomized rectangle for $f$ with expected size $s=1/2^{O(k)}$ and
expected (constant) error $\epsilon\le1/3$. Then there is an approximate majority cover for $f$
having size $2^{O(k)}\cdot n$.
\end{lemma}

The lemma clearly implies $APP(f)\le O(k)+\log n$.
So let us prove the lemma. We independently pick $t=c\cdot (1/s)\cdot n$ rectangles
$R_1,\ldots, R_t$ from the
distribution $D$ for some large enough constant $c$. Our claim is that
this yields the desired approximate majority cover.
Let $w(x,y)$ denote the random variable counting the
number of $R_i$ with label $v(R)\neq f(x,y)$ and $x,y\in R_i$.
Let $h(x,y)$ denote the number of all $R_i$ with $x,y\in R_i$.

Consider the distribution $\mu_{x,y}$ concentrated on $x,y$.
We know that the expected size of a rectangle picked from $D$ is at least
$s$. Since $\mu_{x,y}(R)\in\{0,1\}$, with probability at least $s$
a chosen rectangle contains $x,y$. So $E[h(x,y)]\ge s\cdot c\cdot (1/s)\cdot
n=cn$.

We know $E[w(x,y)]\le \epsilon E[h(x,y)]$, and want to bound
$Prob(w(x,y)\ge 1.1\cdot\epsilon E[h(x,y)])$, which is
maximized if $E[w(x,y)]$ is as large as possible, hence we assume
$E[w(x,y)]=\epsilon E[h(x,y]$.
Using the Chernov bound \begin{eqnarray*}
&&Prob(w(x,y)\ge 1.1\cdot\epsilon E[h(x,y)])\\
&= & Prob(w(x,y)\ge 1.1\cdot E[w(x,y)])\\
&\le& e^{-E[w(x,y)]/300}= e^{-\epsilon E[h(x,y)]/300}\le e^{-\epsilon cn/300}.\end{eqnarray*}
Let $c=O(1/\epsilon)$ be large enough, so that the above probability
is at most $2^{-2n-1}$.
Then the
probability that there exists one of the $2^{2n}$ inputs $x,y$ with
$w(x,y)\ge 1.1\cdot\epsilon
h(x,y)$ is smaller than 1. Consequently there exists a choice of $t$
rectangles so that for all $x,y$: $w(x,y)\le
1.1\cdot\epsilon h(x,y)$.

By Remark~\ref{rem:APPboost} we get an approximate majority cover.
\qed

We now show an exponential gap between (even one-sided) bounded error
uniform threshold covers and approximate
majority covers.

{\bf Proof of Theorem~\ref{the:APPUTsep}.}
Consider the function $BOTH:(\{0,1\}^{2n}\times\{0,1\})\times
(\{0,1\}^{2n})\to\{0,1\}$ defined as follows: \[BOTH((x_1,x_2,a),(y_1,y_2))=
(DISJ(x_1,y_1)\wedge a)\vee(\neg DISJ(x_2,y_2) \wedge \neg a).\]

Hence depending on $a$ the function either computes $DISJ$ on the
first pair of inputs or $\neg DISJ$ on the second pair.

First we show that $APP(BOTH)=O(\log n)$. Note $APP(\neg DISJ)=O(\log
n)$, since $N(\neg DISJ)=O(\log n)$ and $APP(f)\le N(f)$. Hence also
$APP(DISJ)=O(\log n)$, since $APP(f)=APP(\neg f)$ for all $f$.
To find an approximate majority cover for $BOTH$ we take the approximate majority
cover for $DISJ$ and intersect all its rectangles with the rectangle
defined by $a=1$. We also take the approximate majority
cover for $\neg DISJ$ and intersect all its rectangles with the rectangle
defined by $a=0$. The union of these sets of rectangles is an approximate majority
cover for $BOTH$. So $APP(BOTH)=O(\log n)$.

Now we consider $UT^{(1)}_{s,2s}(BOTH)$. We consider a distribution on
inputs in which $a=1$ with probability 1. In this case with probability 1,
$BOTH((x_1,x_2,a),(y_1,y_2))=DISJ(x_1,y_1)$.
Since there is a balanced distribution on inputs so that each
1-rectangle either has size $1/2^{\Omega(n)}$ or error at least
$\epsilon$ for some constant $\epsilon>0$ (see Fact~\ref{fac:Razb}),
we can choose this distribution on the input
positions $x_1,x_2$, and fix $x_2,y_2$ arbitrarily. In this way we get a
balanced distribution with the same properties for $BOTH$ and hence
$UT^{(1)}_{s,2s}(BOTH)=\Omega(n)$ (using Theorem~\ref{the:BDUT}).

Now we consider $UT^{(1)}_{s,2s}(\neg BOTH)$. We may proceed as above, by
fixing $a=0$ and considering the quality of 1-rectangles for $\neg(\neg
DISJ)$. So we get $UT^{(1)}_{s,2s}(\neg BOTH)=\Omega(n)$.
\qed

{\bf Proof of Theorem~\ref{the:APPPP}.}
It is easy to construct a majority cover for $MAJ$. The cover contains
$n$ 1-rectangles defined by
$x_i\wedge y_i$ plus $\lceil n/2\rceil$ 0-rectangles covering $\dom$.
If we have $MAJ(x,y)=1$, then at least
$\lceil n/2\rceil$ 1-rectangles $x_i\wedge y_i$ contain $x,y$, else at most
$\lceil n/2\rceil -1$ 1-rectangles contain $x,y$.

For the lower bound we have to argue that there is a balanced
distribution for which all $1-\epsilon$-correct rectangles have size
at most $1/2^{\Omega(n)}$. Let $n'=6k+2$ be the input
length for some $k$ satisfying $k\equiv
1\mbox{ mod } 2$ and $k\equiv 1\mbox{ mod } 3$.

First we fix $2k$ variables
$x_i,y_i=1$. There are $n=4k+2$ remaining variables. $MAJ(x,y)=1\iff
\sum_{i=1}^n x_i\wedge y_i\ge k+1$ under this fixing. We pretend in the
following that there are $n$ variables.

Let us define the distribution.
Let $\mu_r$ be the uniform distribution on $\{(x,y):|x|=|y|=n/2, |x\cap
y|=k\}$, and let $\mu_a$ be
the uniform distribution on $\{(x,y):|x|=|y|=n/2, |x\cap
y|=k+1\}$. Then let $\mu$ be defined by
$\mu(x,y)=(3/4)\cdot\mu_r(x,y)+(1/4)\cdot \mu_a(x,y)$.
The distribution is obviously balanced.

We have to show that there are no large rectangles with small error,
neither 1-rectangles nor 0-rectangles. This is handled in the
following way.

\begin{claim}
If there is a 1-rectangle of size $\Omega(s)$ and error $O(\delta)$ according to
$\mu$ and $MAJ$, then there is a 0-rectangle of size $\Omega(s)$ and
error $O(\delta)$ according to
$\mu$ and $MAJ$, and vice versa.
\end{claim}

{\bf Proof of the claim.} Assume there is a size $s$ 1-rectangle $R=A\times B$ with
a fraction of $(1-\delta)s$ 1-inputs and $\delta s$ 0-inputs
on $\mu$. Let $\neg
A=\{x:\overline{x}\in A\}$. We claim that $\neg A\times B$ is an
$O(\delta)$-error size $\Omega(s)$ 0-rectangle.
Note that $f(x,y)=0\iff |x\cap y|=k$ and $f(x,y)=1\iff |x\cap
y|=k+1$ under $\mu$, and that $|x|,|y|=2k+1$. Hence $|\overline{x}\cap
y|=|y|-|x\cap y|=2k+1-k-f(x,y)$, and so $f(x,y)\neq f(\overline{x},y)$
with probability 1. So the rectangle $\neg A\times B$ has entries with
reversed function value compared to $A\times B$. The claim follows with
$\mu(x,y)=\Theta(\mu(\overline{x},y))$.\qed

We are going to show that each 0-rectangle has size at most
$2^{-\Omega(n)}$ or has error $\epsilon$ for some constant $\epsilon$. 
Then $APP(MAJ)=\Omega(n)$.

We consider the following way to choose inputs according to $\mu$:
First we choose a {\it frame}, namely a partition of $\{1,\ldots,
n\}$ into sets $z_k$ of size $k$, $z_x,z_y$ of size
$(4k+2-k-1)/2=\lceil3k/2\rceil$, and $\{i\}$ of size 1, uniformly
under all such partitions. Then $x$ is chosen to contain all of $z_k$
and with probability 1/2 also $\{i\}$. $x$ is filled up to a size
$2k+1$ set by choosing uniformly elements of $z_x$. $y$ is chosen
similarly, only with the
filling up done from the set $z_y$. Note that this produces the
distribution $\mu$.

Now we fix $z_k$ arbitrarily. Let $\mu_{z_k}$ denote the corresponding
distribution on inputs. Ignoring the variables in $z_k$ the
players choose sets with an intersection size in $\{0,1\}$, i.e.,
they solve the (complement of the) disjointness problem on a specific
distribution.

We employ Fact~\ref{fac:Razb2} at this point.
A technical problem for the application of this fact is that for
$\mu_{z_k}$ subsets of size $k+1$ are chosen from a size $n-k=3k+2$ universe.
To overcome this we may fix arbitrary disjoint subsets
$s_x,s_y\subseteq\{1,\ldots, n\}-z_k$ of
size $l=k/3+2/3$ each. The variables in $s_x$
are set to 1 in $x$ and the variables in $s_y$ are set to 1 in
$y$. After fixing $z_k, s_x, s_y$ an input is chosen as follows. First
$z_x$ and $z_y$ are chosen, under the condition that they include
$s_x$ resp.~$s_y$, hence the remaining size of these is $\lceil
3k/2\rceil-\lceil k/3\rceil$ each. Then $\{i\}$ is chosen and the
frame is complete. Afterwards an input is chosen as before.
Call the resulting distribution $\mu_{z_k,s_x,s_y}$.

The number of remaining nonfixed variables when choosing according to $\mu_{z_k,s_x,s_y}$
is $n''=3k+2-l$. Disregarding the fixed $l$ elements the size of $x$ and
of $y$ is $k+1-l=(2/3)k+1/3=n''/4$.
So disregarding the fixed inputs we have reached the distribution
$\nu$ of Fact~\ref{fac:Razb2}.

Under $\mu$ the weight of any input $x,y$ can be expressed
as the expectation over all
possibilities to fix $z_k$ and to fix $s_x,s_y$ of the weight of the
input under this fixing.
Namely, \[\mu(x,y)=E_{z_k,s_x,s_y}[\mu_{z_k,s_x,s_y}(x,y)].\]
We know from Fact~\ref{fac:Razb2} that for all $z_k,s_x,s_y$: \[\mu_{z_k,s_x,s_y}(MAJ^{-1}(1)
\cap R)\ge 1/5 \cdot \mu_{z_k,s_x,s_y}(R)-2^{-\Omega(n)}.\]
\begin{eqnarray*}
\mbox{Hence also }&&\mu(R\cap MAJ^{-1}(1))\\
&=&
E_{z_k,s_x,s_y}[\mu_{z_k,s_x,s_y}(MAJ^{-1}(1)
\cap R)]\\
&\ge& (1/5) \cdot E_{z_k,s_x,s_y}[\mu_{z_k,s_x,s_y}(R)]-2^{-\Omega(n)}\\
&=&(1/5)\cdot \mu(R)
-2^{-\Omega(n)}.\end{eqnarray*}
Hence any 0-rectangle for $MAJ$ under $\mu$ either has size
$2^{-\Omega(n)}$, or error $1/5$. Due to our previous claim
within constant factors the same holds for 1-rectangles. So the lower
bound $APP(MAJ)=\Omega(n)$ follows.
\qed
\end{appendix} 
\end{document}